# End of Publication? Open access and a new scholarly communication technology

**Sergey Parinov [1,\*] and Victoria Antonova [2]**

[1]  Central Economics and Mathematics Institute of Russian Academy of Sciences; sparinov@cemi.rssi.ru

[2]  National Research University Higher School of Economics, Moscow, Russia; vantonova@hse.ru

\*  Correspondence: sparinov@cemi.rssi.ru; Tel.: +7-916-914-9051

**Abstract:** At this time, developers of research information systems are experimenting with new tools for research outputs usage that can expand the open access to research. These tools allow researchers to record research as annotations, nanopublications or other micro research outputs and link them by scientific relationships. If these micro outputs and relationships are shared by their creators publically, these actions can initiate direct scholarly communication between the creators and the authors of the used research outputs. Such direct communication takes place while researchers are manipulating and organizing their research results, e.g. as manuscripts. Thus, researchers come to communication before the manuscripts become traditional publications. In this paper, we discuss how this pre-publication communication can affect existing research practice. It can have important consequences for the research community like the end of publication as a communication instrument, the higher level of transparency in research, changes for the Open Access movement, academic publishers, peer-reviewing and research assessment systems. We analyze a background that exists in the economics discipline for experiments with the pre-publication communication. We propose a set of experiments with already existed and new tools, which can help with exploring "the end of publication" possible impacts on the research community.

## 1. Introduction

The focus of this paper is on publication as an instrument of scholarly communication. We discuss the "end of publication" as a very possible outcome of the recent innovations in research information systems [1]. Apart from this communication function, publications undoubtedly have some other value for its authors.

We examine some new trends in modern research information systems development related to Open Access (OA) and scholarly communication technology. Neylon wrote about this - "There are huge opportunities starting to open up for more effective research communication. The massive progress towards Open Access is a core part of this." [2]. We conclude that these trends potentially eliminate research publication as an instrument of scholarly communication, because research information systems tends to allow researchers direct communication and cooperation at the pre-publication stage.

In the most general sense, such changes can mean that the "market place of scientific communication", as it was defined about 20 years ago by Roosndaal and Geurts [3], is now under transformation and is losing its dominating position. The "market place" style of communication based on the publications exchange is being gradually replaced by the direct scholarly communication, which is usual for a research laboratory or a project team.

By the direct pre-publication communication, the research community should achieve better cooperation among the scholars and more transparency in research, which became highly demanded during the last years [4], [5]. We expect important benefits for the society. However, there are many problems on the way to realizing these benefits.

Assuming that researchers do not need to use traditional publications for scholarly communication any more, we should rethink the roles of many societal institutions dealing with publications.

Particularly, the scholarly publishers and journals should rethink their business model. The Open Science Initiative Working Group notes that: "… with the explosive growth of information available through computers and the Internet in recent decades, our rapidly changing societal expectations about having free and rapid access to information, the continued emergence of many new research specializations, and the explosive growth of knowledge creation, the scholarly publishing system has reached perhaps the most significant crossroads in its history." [6]. We also should analyze possible changes of peer-reviewing and research assessment systems because they are based on the scholarly publishing system.

In the next section of the paper, we examine how publication-based scholarly communication mechanism works. We discuss some of its weakness and shortcomings. We define the scholarly pre-publication communication. It can serve the research community more efficiently compared with traditional scholarly publishing system. We conclude that this pre-publication communication creates opportunities to improve the level of transparency in research.

The third section of the paper discusses technical details of research outputs usage, which initiate pre-publication communication. We propose to create an open technical platform for experiments with these new tools.

The fourth section is about an existing background for creating of a platform for experiments. Necessary background has emerged in economics in the form of the Research Papers in Economics (RePEc.org) data sharing platform. These data are generated by an active international research community of RePEc users (about 1800 organizations and 50000 researchers) and are presented by the innovative SocioRePEc tools (sociorepec.org). We discuss what experiments can be fulfilled on this platform.

In the conclusion, we provide a brief overview of challenges arising from shifting research community to pre-publication communication and by eliminating some of the functions of research publications. One can expect serious changes in research culture, formal and informal norms and institutions like scholarly publishing system, per-reviewing and research assessment practices, etc.

## 2. Publication as a traditional scholarly communication instrument

Eisen and Vosshal wrote: "Scientists publish for two reasons: to communicate their work to their colleagues, and to get credit for it in hiring, promotion and funding." [7]. The reason "to communicate their work to their colleagues" is obviously the main here, since another - "to get credit" - can work only if the publication has fulfilled its communicative function, i.e. the publication has reached the "colleagues" and they have evaluated it as more or less valuable for their work.

Typically, scientists find research publications through the traditional print-based and peer-review scholarly publishing system (e.g. Springer-SCOPUS) or at the aggregator-type information systems (e.g. RePEc), which are harvesting publications metadata from Institutional Repositories (IR) or other OA sources.

While reading publications they mentally select from the text some useful pieces of information. We can call them the "research artifacts". Usually researchers manipulate with the artifacts to find out scientific relationships among them. They then create their own research artifacts. To share the created relationships and artifacts with the community, researchers have to use the scholarly publishing system or IR. In both cases, researchers have to present the newly created relationships and artifacts in a form of a canonical research manuscript, like working paper, article, book, etc. with citations, reference lists and some other mandatory attributes. They submit manuscripts to publishers or deposit them at IR. After publication, researchers typically are monitoring publication' usage. They may be expecting some occasional comments on it. As shown at Figure 1, researchers permanently repeat this cycle.

Scholarly communication as supported by the traditional scheme at Figure 1 has fundamental weaknesses and shortcomings.

While manipulating artifacts researchers make with them some mental experiments to combine them with their own research results. This is some kind of "trial and fail" action. Because of such "trial and fail" experiments, researchers either accept some artifacts by connecting them with their own research results or reject them. In both cases, researchers eventually get some individual experience about the scientific value and/or the potential impact of these tested research artifacts. This information can be very useful for authors of the tested artifacts and for the community at large. However, currently only a small piece of this information becomes public.

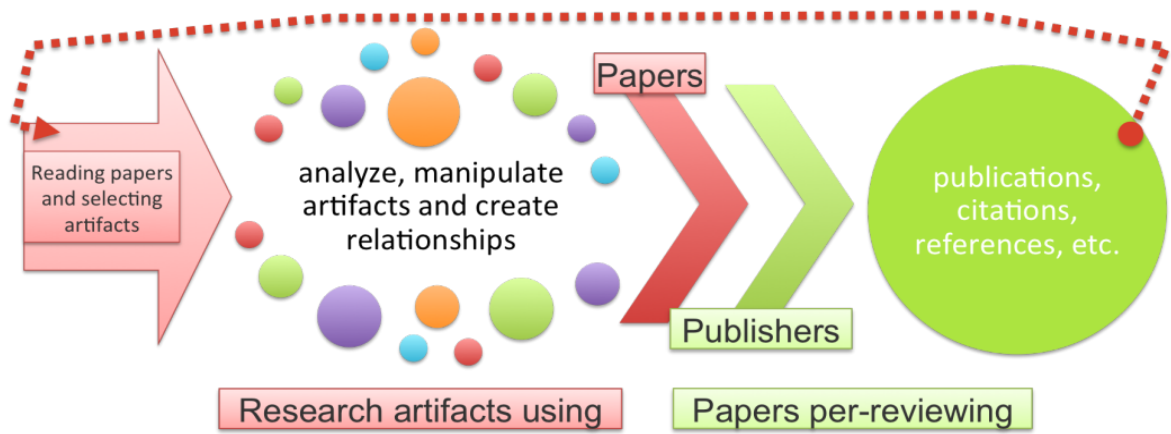

Figure 1. Traditional scholarly communication scheme through academic publishers and journals infrastructure

In the traditional scheme of the global scholarly communication, most authors hardly get any feedback on their papers. They usually do not know who, why and what for tried to use their research results and fails. If someone really used the authors' results and cited them in their own papers, the authors had no chance to assist them with the proper usage of the results, to correct a wrong usage or to get a better effect from using their results. From a point of research cooperation, especially compare with how scholarly communication works in a research laboratory, this situation looks inefficient.

Generally, researchers do not manipulate whole publications but the research artifacts contained in the publications. They also manipulate the relationships between artifacts. Why they could not share these artifacts and relationships as their **micro research outputs** with the other researchers?

In this context, the micro research output is a unit of a research composition, created by a researcher for personal or public research reuse. It can be an artifact with some added value, like a comment, or a scientific relationship.

Let us analyze a hypothetical situation. Scholarly communication is initiated by sharing not publications, but micro research outputs. A research information system notifies authors of research outputs all types if their outputs were used someone to create micro research outputs. In this case, scholarly communication appears in the research community at the stage, which comes before the researchers publish their papers. We can call this the **"pre-publication"** communication.

When the pre-publication communication takes place, researchers may no longer need to formatting their micro research outputs as papers and making them published in order: (a) to distribute their research results globally; (b) to get feedback on them; (c) to have them used by other researchers in producing new research results; (d) to cooperate with researchers who use their results.

The pre-publication communication is not principally new for the research community. It is used by researchers for communicating within a laboratory, within a project team, etc. If the global scholarly communication that is based now on publications using the scholarly publishing system could be performed in a way similar to communications in a laboratory, the community gets multiple benefits, such as well known strong characteristics of the laboratory-style research. First, it is the better transparency in research, which is highly demanded now by the community at large [4], [5].

The main challenge of improving transparency in research is a public visualization of the bigger part of the process of the researcher's scientific analysis and creativity. At the Figure 1, it is the area "analyze, manipulate artifacts and create relationships". For scholarly communications based on publications, the essential part of researcher's creativity is traditionally hidden from the research community, because it is carried out in the mind of the researcher.

Suitable pre-publication communication technology can generate public data that will make research more transparent. The obvious question here is how the pre-publication communication technology should work in the global scale. To meet the challenge it should be designed to replace effectively the publication-based communication supported by the academic publishing system.

### 3. Pre-publication communication technology

In general, a pre-publication communication technology should be a part of a research information system (RIS). This RIS should allow researchers to make actions, which can be recognized by RIS as a reason for initiating the pre-publication communication. Researchers make these actions when they express publicly: (1) what kind of research results (artifacts) they select as interesting; (2) which ones they use in their work; and (3) how these results have been used in their study. Tracing the actions, RIS initiates pre-publication communication when one researcher used/cited in his/her own research outputs some research result belonged to another researcher. In this way, the pre-publication communication connects the two groups of researchers – those, who used someone's research results; and the authors of the research outputs, contained these used results.

Taking into account the existing RIS background, this general scheme can be implemented in the following way.

While reading a paper's abstract at the metadata web page or the full text in PDF file, a RIS user can select text fragments. Thus, the fragment can be selected from either the abstract or the full text. After selecting some text fragment, a user sees a form on a screen. From this form, a user can create different types of what we call the micro research outputs:

1. A user's comment to the selected fragment. Such comments will be visible to the next reader. A similar user feature already exists in some research information systems (e.g. at ResearchGate.net). There is some infrastructure to aggregate and exchange them (e.g. at Hypothes.is).

2. A research assertion (the same as at Nanopub.org) that a user concludes from the selected fragment. Technically, the use of this feature implies a creation of a nanopublication[1], which can be integrated into appropriated infrastructure [8].

3. A quotation that a user has decided to select from the paper and share with the community as a research artifact. A quotation is supplemented by a user's comment explaining why it has been selected and shared. By making such quotations users can prepare some text fragments to create scientific relationships between them.

4. A micro research paper where the user's comment is the primary content and the selected fragment is just a base for it.

Additionally to these four research usage actions, there is also a fifth option, which is available for RIS users in some other way. It is the creation of a scientific relationship between any pair of micro and/or traditional research outputs, available at the RIS content [9]. It allows citing any piece of a research output. If researchers discovers interdependences or relations between any two pieces of research content, they can express these by making scientific relationships. Kogalovsky and Parinov [10] present an initial taxonomy of scientific relationships.

All these research usage actions generate private or public micro outputs that are being stored in the RIS. These micro outputs are automatically linked with the initial paper and with the personal profile of its creator. While reading a paper at RIS content, a researcher can see all created research micro outputs linked with the particular paper.

Therefore, pre-publication communication is being initiated immediately after any of these 5 research usage actions have been made. The author of the used/cited research output receives e-mail notification with data about what the usage/citation means. She/he can react to these usage/citation actions.

The full scheme of pre-publication communication on behalf of an author of some research output, which is used by someone, is illustrated at Figure 2. It works as follows [11]:

1. The author of publication or micro output receives signals (email notification) from the RIS when someone (a user) used his/her research results. This starts the communication between the author and the user.

2. When the author receives this signal, she/he can assist the user in the proper use of the research results. Thus, there may be an act of cooperation between the author and the user.

3. The author can modify the research outputs to enhance the effect of the research results use. This is also an act of cooperation.

---

1 http://nanopub.org/wordpress/?page_id=65

4. The communication between the author and the user may be published in the scientific information space. In this case, the public nature of the exchange may generate competition between different authors to serve the users.

5. All data are stored in the RIS. They form a public portrait of the researcher. The portrait includes all actions taken and the reactions that allow evaluating scientific reputation of the researcher. Correspondingly, this significantly increases the responsibility of the researcher for her/his actions in comparison to the situation created by the traditional mechanism of communication.

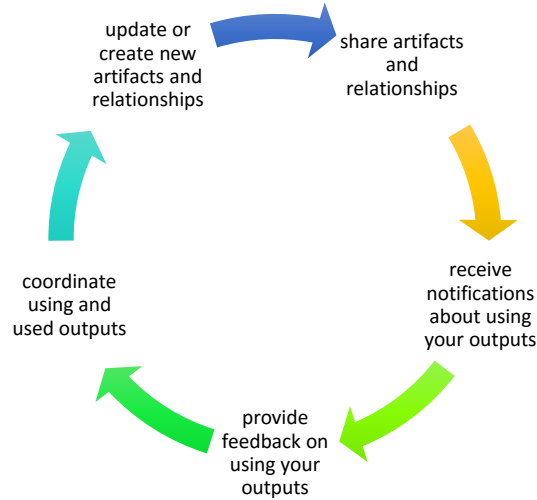

Figure 2. Pre-publication communication scheme

Pre-publication communication based on such facilities allows researchers to put smaller pieces of their work into scientific circulation and re-use. Thus, we can implement Neylon's idea that any paper is an aggregation of objects. "If we take this view of objects and aggregates that cite each other, and we provide details of what the citations mean (this was used in that, this process created that output, this paper is cited as an input to that one) then we are building the semantic web as a by-product of what we want to do anyway" [12].

Thinking about practical steps to put scholarly pre-publication communication in real life, we should take into account Neylon's notes [13]: "So in short, publish fragments, comprehensively and rapidly. Weave those into a wider web of research communication, and from time to time put in the larger effort required to tell a more comprehensive story. This requires tools that are hard to build, standards that are hard to agree, and cultural change that at times seems like spitting into a hurricane. Progress is being made, in many places and in many ways, but how can we take this forward today?"

We agree with the authors of the Publons research information system, when they wrote about innovative scholarly communication tools: "It is hard to say exactly what these new tools will be. Their development will require a great deal of experiment and iteration. Therefore, instead of attempting to a priori identify and build the right tool, we need to develop a framework (and culture) in which experimentation with different forms of collaboration is encouraged." [14].

Thus the right first step towards including the pre-publication communication into research practice is to create a public RIS as an open framework or a platform for experiments with tools, built use cases, etc.

## 4. A platform for experiments with pre-publication communication

Research Papers in Economics (RePEc) is a data sharing platform. It unities an active international research community of RePEc users. RePEc works as an aggregator-system collecting metadata of about 2 million research outputs from over 1800 research organizations worldwide. It also includes personal profiles of about 50000 researchers who linked their profiles with their available at RePEc research outputs and with profiles of their affiliations.

Using previously tested approaches [11] and newly created technology [15] we have designed a research information system called SocioRePEc, at https://sociorepec.org/. It is based on RePEc data. It is an open non-commercial project funded by grants and donations. SocioRePEc's main added value is to allow for experiments with advanced research outputs usage and some initial pre-publication communication. The SocioRePEc team is seeking to attract RePEc users and research organization for further development of these tools.

SocioRePEc tools allow for different kind of experiments with pre-publication communication but there is a problem of incentives to do the experiments. As Publons authors wrote - "… in order to build innovative systems for collaboration we need to modify this incentive structure. A corollary of this is that any new approach must augment the current system, at least initially" [14].

To create incentives we propose improving global scholarly communication by embedding the SocioRePEc tools into the traditional communication scheme. See an illustration at Figure 3.

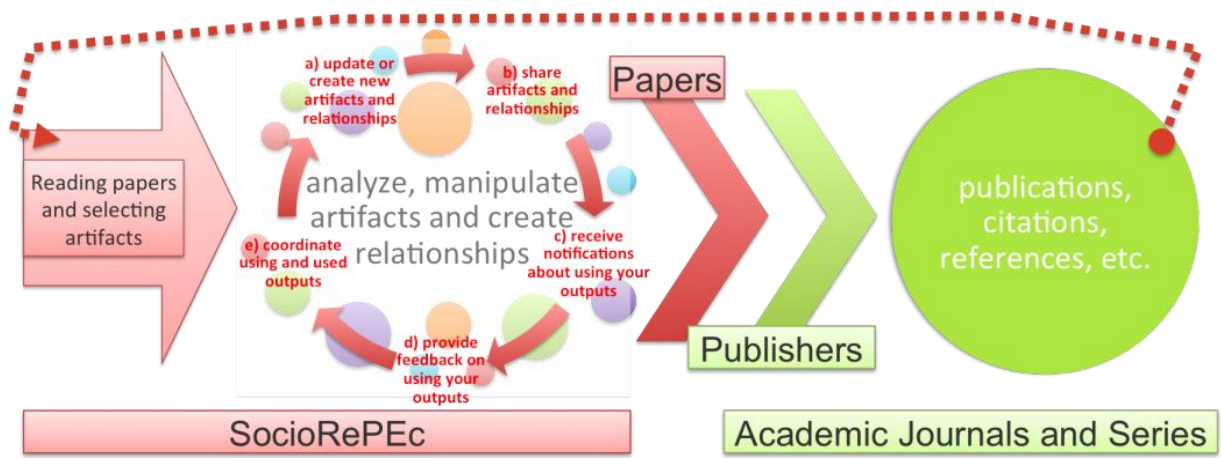

Figure 3. Scholarly communication scheme as a combination of SocioRePEc facilities and academic journals/publishers infrastructure

This combination with SociRePEc can augment the traditional academic publishing system. All traditionally published papers and papers from OA sources collected by RePEc are available at SocioRePEc for research usage actions listed in the previous section. SocioRePEc services are tracing the usage actions and are notifying authors of the used research outputs. Notified authors can react to the usage actions made with their research outputs. As a result, researchers will produce better papers and submit them for publishing in traditional ways.

A combination of these SocioRePEc facilities with the existing academic journals/publishers infrastructure could create some new opportunities for the research community.

It could provide researchers with more efficient mechanism of the global scholarly communication as compared to the existing academic publishing system. Thus new communication mechanism should co-exist with the formal publishing infrastructure as its informal complement.

As Pearce et al. underline, "new web based technologies are then a necessary, but not sufficient, condition for a radical opening up of scholarly practice. In this sense digital scholarship is more than just using information and communication technologies to research, teach and collaborate, but it is embracing the open values, ideology and potential of technologies <…> to benefit both the academy and society. Digital scholarship can only have meaning if it marks a radical break in scholarship practices brought about through the possibilities enabled in new technologies. This break would encompass a more open form of scholarship» [16] (p .40-41).

Thinking about testing the pre-publication communication usefulness, we propose some experiments with SocioRePEc facilities. They can help with exploring the end of publication consequences.

1. Competitive selection. The basic pre-publication communication provided by SocioRePEc is public. That means the system allows experiments with creating additional competition. Other researchers can trace the "author"<->"user" pre-publication communication and can compete with the author by offering the user a better research result or more efficient solution for her/his "demand".

2. Identification of the "neighbors". We can think of researchers using research outputs of other researchers as "neighbors" in the global scientific labor division system. Pre-publication communication can help researchers to find out who their neighbors are. It will initiate scholarly communication between the neighbors to get them better collective intelligence and additional benefits from their direct research cooperation.

3. Stronger research coordination. Pre-publication communication can initiate coordination between a researcher who created a research result and another researcher who used it. They can interactively adjust and adapt their "supply" and "demand" to get better mutual effect from their research cooperation.

4. Threats of the "end of publication". Do researchers appreciate that pre-publication communication is an instrument for identifying problems in and reducing potential threats to the credibility of their work? This question requires also some additional qualitative study on how the research culture (formal and informal norms, rules and motivation) be developed that can lead researchers to adoption of the pre-publication scholarly communication practice.

5. Publication as aggregation. It is also important to find out what could motivate scholars to adopt the idea that the future of research publication is aggregation. Neylon wrote about this: "For me, the "paper" of the future has to encompass much more than just the narrative descriptions of processed results we have today. It needs to support a much more diverse range of publication types, data, software, processes, protocols, and ideas, as well provide a rich and interactive means of diving into the detail where the user in interested and skimming over the surface where they are not. It needs to provide re-visualisation and streaming under the users control and crucially it needs to provide the ability to repackage the content for new purposes; education, public engagement, even main stream media reporting." [12].Possible questions for the experiments are: What kind of forms in general can research outputs usage have, e.g. in economics? Will researchers agree to share micro research outputs in order to benefit from the pre-publication communication? Under what circumstances could researchers adopt the idea of "publication as aggregation"?

6. Transparency in research. What changes in research practice can initiate global pre-publication scholarly communication between authors and users of research outputs? How this can improve transparency and credibility of their research findings? Answering these questions will imply some study of, for example, the community of RePEc users. We see them as a pro-active group of scholars open to innovations in the field of global scholarly communication technology.

7. Open scholar challenges. Developing scholarly communication in a direction of pre-publication communication and sharing micro research outputs, scholars become more open. An open scholar "is someone who makes their intellectual projects and processes digitally visible and who invites and encourages ongoing criticism of their work and secondary uses of any or all parts of it--at any stage of its development" [17]. We agree with Pearce et al., that "this is a significant and challenging step for scholars, especially when faced with norms and values that oppose, hinder, or fail to recognize these forms of scholarship". [16] (p. 41).

## 5. Conclusions

We think that current development trends of OA and global scholarly communication technology allow us to talk about the "end of publication" as a communication instrument in a long-term perspective. We are expecting different social and institutional impacts generated by shifting research communities to pre-publication communication and by eliminating some of the functions of research publications. It could result serious changes in research culture, formal and informal norms and institutions like traditional scholarly publishing system and related peer-reviewing and research assessment practices.

Benefits from pre-publication communications can also challenge the OA movement. It can come to some kind of crises, because researchers can start sharing all their research results as micro outputs and scientific relationships before they publish them as papers. Under this situation, access to papers still can be restricted, but all research results from these papers are open and publically available at some RIS.

At the same time pre-publication scholarly communication reproduces in a stronger way some traditional features, e.g. as those described 20 years ago by Roosendaal and Geurts - "…authors want to publish more and have their product widely available, while readers want to read less, but want to be informed of all that is relevant for their research at hand. They want this information available just in time. They want to be guaranteed that they can and will be informed of all that is relevant to them." [3]

**Acknowledgments:** We are grateful to Thomas Krichel for his useful comments. The Russian Foundation for Basic Research funded this study in a part of the global scholarly communication technology, grant 15-07-01294-a.

## References

1. Kramer, B. and Bosman J. Innovations in scholarly communication - global survey on research tool usage. F1000Research 2016, **5**:692 (doi: 10.12688/f1000research.8414.1)

2. Neylon C. Architecting the Future of Research Communication: Building the Models and Analytics for an Open Access Future. *PLoS Biol*. **2013**, *11(10)*: e1001691. doi:10.1371/journal.pbio.1001691

3. Roosendaal H. E., Geurts P. A. T. M. Forces and functions in scientific communication: an analysis of their interplay. – **1997**., http://doc.utwente.nl/60395/1/Roosendaal97forces.pdf

4. Fanelli, D. How Many Scientists Fabricate and Falsify Research? A Systematic Review and Meta-Analysis of Survey Data. *PLoS ONE* **2009**, *4* (*5*): e5738. doi:10.1371/journal.pone.0005738

5. Chang, A. C.; Li P. Is Economics Research Replicable? Sixty Published Papers from Thirteen Journals Say "Usually Not". *Finance and Economics Discussion Series* **2015**-083. Washington: Board of Governors of the Federal Reserve System. doi:10.17016/FEDS.2015.083

6. Open Science Initiative Working Group. Mapping the Future of Scholarly Publishing", 1st edition // Seattle: National Science Communication Institute, January 2015. Available online: http://nationalscience.org/wp-content/uploads/2015/02/OSI-report-Feb-2015.pdf (accessed on 27 March 2016)

7. Eisen, M.; Vosshall, L.B. Coupling Pre-Prints and Post-Publication Peer Review for Fast, Cheap, Fair, and Effective Science Publishing. Blog post. Published: January 21, 2016. Available online: http://www.michaeleisen.org/blog/?p=1820 (accessed on 27 March 2016)

8. Kuhn, T.; Chichester, C.; Krauthammer, M.; N´uria Queralt-Rosinach; Verborgh, R.; Giannakopoulos, G.; Axel-Cyrille Ngonga Ngomo; Viglianti, R.; Dumontier, M. Decentralized Provenance-Aware Publishing with Nanopublications. *Peerj*, **2016**. Available online: https://peerj.com/preprints/1760.pdf (accessed on 27 March 2016)

9. Parinov, S. Semantic enrichment of research outputs metadata: new CRIS facilities for authors. In *Metadata and Semantics Research Communications in Computer and Information Science* **2014**, *Volume 478*, pp 206-217, Available online: http://link.springer.com/chapter/10.1007/978-3-319-13674-5_20 (accessed on 27 March 2016)

10. Kogalovsky, M., Parinov, S. The taxonomy of semantic linkages of information objects in research digital library content. *Automatic Documentation and Mathematical Linguistics*, September **2015**, Volume 49, Issue 5, pp 163-171, Available online: http://link.springer.com/article/10.3103/S0005105515050027 (accessed on 27 March 2016)

11. Kogalovsky, M., Parinov, S. Scholarly communication in a semantically enrichable research information system with embedded taxonomy of scientific relationships. In "*Knowledge Engineering and Semantic Web*". Editors: Klinov, Pavel, Mouromtsev, Dmitry (Eds.). Springer International Publishing, **2015**, Volume 518, pp 87-101, Available online: https://socionet.ru/publication.xml?h=RePEc:rus:mqijxk:38&l=en (accessed on 27 March 2016)

12. Neylon, C. The future of research communication is aggregation, Science in the Open Blog, Published: 10 APRIL 2010. Available online: http://cameronneylon.net/blog/the-future-of-research-communication-is-aggregation/ (accessed on 27 March 2016)

13. Neylon C. What would scholarly communications look like if we invented it today? Science in the Open Blog, Published: 2 SEPTEMBER 2010. Available online: http://cameronneylon.net/blog/what-would-scholarly-communications-look-like-if-we-invented-it-today/ (accessed on 27 March 2016)

14. Preston A, Johnston D. The Future of Academic Research. figshare. Available online: https://dx.doi.org/10.6084/m9.figshare.871466.v1 2013.

15. Parinov, S.; Lyapunov, V.; Puzyrev, R.; Kogalovsly, M. Semantically Enrichable Research Information System SocioNet. In the *Knowledge Engineering and Semantic Web*. Editors: Klinov, P., Mouromtsev, D. (Eds.) Springer International Publishing, **2015**, *Volume 518*, pp 147-157. Available online: http://link.springer.com/chapter/10.1007%2F978-3-319-24543-0_11 (accessed on 27 March 2016)

16. Pearce, N., Weller, N., Scanlon, E. & Ashleigh, M. Digital Scholarship Considered: How New Technologies Could Transform Academic Work // *In Education*. **2010**, 16(1), pp. 33-43.

17. Burton, G. The open scholar. Academic Evolution. **2009**. Available online: http://www.academicevolution.com/2009/08/the-open-scholar.html (accessed on 15 April 2016)